# Reconstruction of Angstrom resolution exit-waves by the application of drift-corrected phase-shifting off-axis electron holography

J. Lindner[1] and U. Ross[1], T. Meyer[1], V. Boureau[3], M. Seibt[4] and Ch. Jooss[1,2]

[1] Institute of Materials Physics, University of Goettingen, Friedrich-Hund-Platz 1, 37077 Goettingen

[2] International Center for Advanced Studies of Energy Conversion (ICASEC), University of Goettingen, D-37077 Goettingen, Germany

[3] Interdisciplinary Center for Electron Microscopy, École Polytechnique Fédérale de Lausanne, CH-1015 Lausanne, Switzerland

[4] 4th Institute of Physics – Solids and Nanostructures, University of Goettingen, Friedrich-Hund-Platz 1, 37077 Goettingen

## Abstract

Phase-shifting electron holography is an excellent method to reveal electron wave phase information with very high phase sensitivity over a large range of spatial frequencies. It circumvents the limiting trade-off between fringe spacing and visibility of standard off-axis holography. Previous implementations have been limited by the independent drift of biprism and sample. We demonstrate here an advanced drift correction scheme for the hologram series that exploits the presence of an interface of the TEM specimen to the vacuum area in the hologram. It allows to obtain reliable phase information up to $2\pi/452$ at the 1 Å information limit of the Titan 80-300 kV environmental transmission electron microscope used, by applying a moderate voltage of 250 V to a single biprism for a fringe spacing of 1 Å. The obtained phase and amplitude information is validated at a thin Pt sample by use of multislice image simulation with the frozen lattice approximation and shows excellent agreement. The presented method is applicable in any TEM equipped with at least one electron biprism and thus enables achieving high resolution off-axis holography in various instruments including those for *in-situ* applications. A software implementation for the acquisition, calibration and reconstruction is provided.

## 1   Introduction

Transmission electron off-axis holography is a phase retrieval technique, which enables access to the full complex exit-wave of thin samples [1]. The additional spatially resolved phase information has been used to reveal physical properties such as electric field distribution around dislocations in thin film solar cells [2], grain boundaries [3], mean inner potentials [4–6], quantum wells [7] and enabled biased semiconductor analysis [8,9]. Other applications are e.g. the measurement of magnetic flux in order to study magnetic domain structures in electron microscopy [10,11], elastic strain measurements [12], as well as to automatically correct residual aberrations [13,14]. For a recent review, see [15].

One major challenge in the determination of potential distributions at material interfaces lies in the different length scales covered by the slowly varying fields of space charge layers versus the contribution of atomic potentials. The latter varies with the lattice periodicity and tends to be of much higher magnitude. Hence, a suitable quantification method must be simultaneously sensitive for low and high spatial frequencies in order to be able to separate both contributions. One of the first attempts to separate the atomic and the surface potentials of a solid [16] additionally showed that microscope aberrations need to be corrected to reveal quantitative information. An important motivation for the study of interface potentials comes from electrochemistry, where *in-situ* environmental transmission electron microscopy allows accessing solid-liquid interfaces at atomic scales [17,18]. Therein the local potential is assumed to play a major role for the reaction mechanisms [19]. Here, full spatial frequency domain sensitivity becomes crucial to resolve the potential contributions of the last atomic column as well as that of space charge regions.

The different types of electron holography techniques and their spatial frequency domain limitations are summarized in short in the following. In off-axis holography [20,21], an interference pattern (hologram) is formed by an electron biprism. The complex exit wave is reconstructed by a digital aperture in Fourier space around one of the sidebands and usually corrected by an empty reference hologram. While off-axis holography provides quantitative phase information, the aperture radius, which is usually 1/3 to 1/2 of the distance of centerband to sideband [22], hampers the reconstruction of higher spatial frequencies thus limiting the spatial resolution. Increasing spatial resolution, resp. lowering fringe spacing via higher biprism voltage leads to a competing decrease of the hologram visibility. The underlying reason is that for a narrower fringe spacing the contrast of neighbouring fringes can be reduced by the point spread function of the camera, thus reducing the visibility. Furthermore, a higher biprism voltage reduces the visibility due to spatial incoherencies of the electron source. Nevertheless, atomic resolution was achieved under optimized experimental conditions, using a high brightness electron gun, careful aberration correction and accumulated image contrast at high camera exposure times [23].

In contrast, in-line holography [24,25] reconstructs the reference wave by acquiring a focal series and employing maximum likelihood algorithms [26], but the lowest spatial frequency of the reconstruction is limited by the defocus variation [27]. A comprehensive comparison of both electron holography techniques can be found in [28]. A discussion of the detectable specific quantum states and their limitations due to the different reconstruction techniques is given by [29]. In order to achieve quantitative phase reconstruction over the complete frequency band, hybrid electron holography has been developed as an attempt to combine in-line and off-axis electron holography [30,31]. Here, the wave function obtained by off-axis holography was used as initial guess for the maximum likelihood reconstruction, which is experimentally challenging since the experimental conditions of the two types of holography experiments must be kept identical.

An alternative approach aims to enhance the spatial resolution of off-axis holography by increasing the sideband aperture and reducing the problem of overlap of side- and centerband Fourier amplitudes. This can be achieved by π-phase-shifting holography. It is a double exposure technique, that records two holograms with a relative phase shift of π induced by a small beam tilt. The subtraction of the holograms substantially reduces the centerband signal, and thus increases the possible sideband aperture radius during the conventional Fourier reconstruction up to the distance between center- and sideband in simulations [32]. In practise, a resolution of one nanometer has been achieved with this method [33]. A very promising technique to avoid the Fourier operations in the reconstruction process entirely has been proposed by Ru et al. [34] in the form of phase-shifting electron holography (PS-EH). This method circumvents the low-pass filtering that happens in the conventional the Fourier space reconstruction and as a consequence its spatial resolution is not connected to the fringe spacing of the hologram. This has the important advantage of using lower biprism voltages, i.e. larger fringe spacing and resulting increased fringe contrast, yielding an improved signal to noise ratio of the reconstruction.

In contrast to medium resolution holography, atomic resolution holography requires to take into account the influence of biprism- and sample drift in the recorded holograms. This makes a correction unavoidable. Moreover, the use of medium biprism voltages shrinks the total hologram width compared to high potentials. Consequently, parts with stronger Fresnel modulations are included in the recorded holograms. These increased Fresnel modulations have to be compensated to determine the specimen drift by cross-correlation schemes, since they lead to undesirable perturbation of signals in the Fourier images. One pathway to eliminate Fresnel modulations in phase-shifting holography is the application of multiple biprisms [35], which strongly improved the phase resolution and revealed the local behaviour of space charges at the p–n junction under different biasing conditions.

In this article, we explore the requirements to apply phase-shifting electron holography in electron microscopes dedicated to *in situ* studies and hence typically equipped with a single biprism with the goal to establish object wave reconstruction over the complete spatial frequency domain up to atomic resolution. We introduce a robust drift correction scheme by using empty reference holograms to reduce the Fresnel contributions in the sample hologram, allowing to apply well-established cross-correlation based drift correction as the essential step for reaching atomic resolution. The proposed concept is demonstrated by using experimental data obtained from a thin platinum single crystal and validated by the comparison of the reconstructed object wave to simulated multislice data.

In the following, we will first introduce the general concept of phase-shifting holography and the reference correction in the ideal, drift-free case in section 2. Section 3 discusses the influence of specimen and biprism drift. Then, we present the procedure developed to correct sample drift in the phase-shifted hologram series. The experimental details are given in section 4. In section 5, the drift correction is demonstrated by the PS-EH reconstruction of an ultrathin crystalline platinum foil in [110] zone axis at high magnification. After reconstruction of the image-wave, residual first order lens aberrations are corrected and the resulting phase and amplitude information is compared to frozen lattice multislice simulations. The comparison demonstrates excellent match in the phase, while the amplitude is slightly reduced due to inaccuracies that are not fully captured in the simulation. Section 6 discusses the results and Section 7 closes with a short summary.

## 2    Phase-shifting holography

### 2.1    Concept of phase-shifting electron holography

The concept of phase-shifting holography, as first developed by Ru et al. [34], is introduced in Figure 1. The technique consists of recording a series of off-axis holograms at different beam tilts. This creates a series of phase-shifted holograms, each of them having a phase shift $\varphi_n$ constant for all positions. We name $\varphi_n$ as the initial phase shift controlled by the beam tilt.

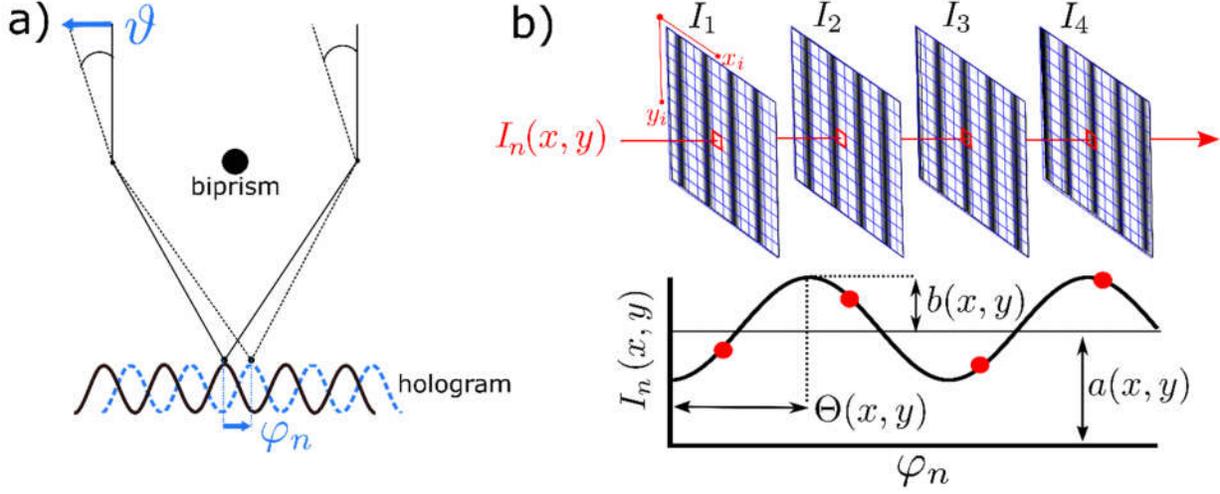

**Figure 1:** *a) Tilting the electron beam by $\vartheta_n$ leads to a position shift of the hologram fringes. Since the intensity of the interference fringes in the ideal hologram follows a cosine, a phase shift $\varphi_n$ can be identified. b) Concept of pixel projection through the phase-shifted hologram series $I_n$. With the knowledge of the initial phase shift $\varphi_n$, the intensity series of each pixel can be subjected to a least-square fit to obtain a(x,y), b(x,y) and $\Theta(x, y)$, from which the image wave can be calculated.*

The related intensity changes of the n-th hologram of the series at the position x and y is described by

$$I_n(x,y) = |\Psi_{img,n}(x,y) + \Psi_{ref,n}(x,y)|^2 \qquad (1)$$

$$I_n(x,y) = \underbrace{A_{ref}(x,y)^2 + A(x,y)^2 + I_{in}}_{a(x,y)} + \underbrace{2\mu(x,y)\ A_{ref}(x,y)\ A(x,y)}_{b(x,y)} \cos\underbrace{\left[2\pi(q_{cx}\ x + q_{cy}\ y) + \Phi(x,y) + \varphi_n\right]}_{\Theta(x,y)} \qquad (2)$$

where, $\Psi_{img,n}$ describes the image wave, $\Psi_{ref,n}$ the reference wave, $A(x,y)$ is the image wave amplitude, $A_{ref}(x,y)$ the reference wave amplitude and $\Phi(x,y)$ the phase of the image wave to be measured, $\mu(x,y)$ describes the local visibility (fringe contrast) and $I_{in}$ describes incoherent intensity contributions for completeness. For all other illumination conditions held constant, the related intensity change at each position follows a cosine function, as shown in eqs. (1) and (2) and Figure 1. Hence, for each pixel of the camera, the offset $a(x,y)$, the amplitude $b(x,y)$ and the reconstructed phase $\Theta(x, y)$ of the cosine function (Figure 1 b)) can be extracted from such tilt series. This is applicable, since the fringe spacings $T_x = 1/q_{cx}$ and $T_y = 1/q_{cy}$ in x- and y-direction, respectively, do not change as long as biprism voltage and orientation are kept constant. The carrier frequency $\bar{q}_c = (q_{cx}, q_{cy})$ adds a constant phase ramp to the reconstructed phase, $2\pi(\bar{q}_c\ r)$, which can be measured from any vacuum region of a hologram to recover the phase of the image wave $\Phi(x,y)$. We emphasize here, that Eqs. (1) and (2) describe the hologram in the coordinate system $r = (x,y)$ fixed to the camera. The reconstruction of $I_n(x,y)$ is done by linear fitting of the cosine function for fixed $(x,y)$ location to determine the three images $a(x,y)$, $b(x,y)$ and $\Theta(x,y)$ by Matrix inversion as proposed in [34].

In order to apply the fitting procedure of Eq. (2), the initial phase shifts $\varphi_n$ must be determined. This can be done *a posteriori* at a vacuum region of a recorded hologram. The initial phase shifts of a tilt series should sufficiently sample the cosine for the fit by being dispersed in the $[0,2\pi]$ range and thus ensure a well-conditioned matrix inversion. In practice, $\varphi_n$ is determined by extracting the phase shift at the carrier frequency $\bar{q}_c$ according to

$$\tilde{I}_n(q) = T(I_n(x,y)) \qquad (3)$$

$$\varphi_n = \tan^{-1}\left(Im\left[\tilde{I}_n(q_{cx}, q_{cy})\right] / Re\left[\tilde{I}_n(q_{cx}, q_{cy})\right]\right). \qquad (4)$$

This allows to quantify the phase shift $\varphi_n$ between each member of the series $I_n(x,y)$. In order to realize a series of phase shifts $\varphi_n$ with beam tilt, its direction and magnitude need to be calibrated with respect to the voltage and orientation of the biprism. Experimentally, at 300 keV primary energy, a beam tilt of a few μrad results in a phase shift of $\approx 2\pi$ for typical biprism voltages of around 200-300 V and can be realized by the gun tilt deflector system of the microscope. Details of the calibration procedure are shown in the SI section 1 or in the GitHub repository [36].

Due to Fresnel diffraction at the biprism edges, deviations of the intensity from the ideal cosine function in Eq. (2) occur, which reduce the fit accuracy slightly. In principle, the cosine could be replaced by finite Fourier series [37] to account for the Frensel modulations. However, a single cosine is used in this work, since the gain in fit accuracy is small compared to the possible error introduced by overfitting. More details can be found in the SI section 2.

## 2.2 Reference correction

Similar to conventional off-axis holography it is advantageous to acquire an empty reference outside the sample area, in addition to $I_n(x,y)$. In this case, the empty reference is a phase shifted series $I_{n,ref}(x,y)$, that can be used to correct distortions of the electron optical system during the reconstruction process. This section assumes, that the specimen drift is absent or already corrected as described in Section 3. The reference correction begins with the two hologram series:

$$I_n(x,y) = a(x,y) + b(x,y) \cos(\Theta(x,y) + \varphi_n) \qquad (5)$$

$$I_{n,ref}(x,y) = a_{ref}(x,y) + b_{ref}(x,y) \cos(\Theta_{ref}(x,y) + \varphi_{n,ref}) \qquad (6)$$

After reconstruction of both series by linear fitting of each pixel series, the reference correction of the amplitude $A(x,y)$ and the phase $\Phi(x,y)$ of the exit wave can be applied. Equation (2) gives:

$$\frac{A(x,y)}{A_{ref}(x,y)} = \sqrt{2\frac{a(x,y)}{a_{ref}(x,y)} - 1} \quad (7)$$

$$\Phi(x,y) = \Theta(x,y) - \Theta_{ref}(x,y) \quad (8)$$

Note, that the incoherent intensity contribution $I_{in}$ of a(x,y) has been neglected. This approximation thus holds if $I_{in} \ll A_{ref}^2$. Alternatively, the amplitude can be calculated by division with the b(x,y)-parameters at the cost of the possible error from different local visibilities µ(x,y) of reference and specimen holograms [38]. Due to the large difference in signal to noise level, the approach in Eq. (7) is preferable compared to the determination from b(x,y). The phase $\Phi(x,y)$ is determined except for a constant, which can be obtained by evaluation of the mean phase of the vacuum region, which should be zero by convention.

## 3 Drift correction

### 3.1 Drift influence

Any mechanical drift of the biprism induces an offset on the phase of the holograms which will add to the initial phase shift created by the beam tilt (see SI section 1). For moderate biprism drift $< 2\pi$ this change of the phase can be considered as being approximately included in the determination of the initial phase $\varphi_n$ by Eq. (4), due to the a posteriori determination from the recorded hologram.

In order to reach atomic resolution by use of phase-shifting holography, an accurate and robust specimen drift correction is essential due to the extraction of the phase $\Phi(x,y)$ from individual fits at each image position $(x,y)$. Indeed, sample drift is critical at atomic resolution and needs to be corrected to an accuracy significantly better than the desired spatial resolution of the measurement, 1 Å in this study. An uncorrected specimen drift leads to an intermixing of amplitude and phase information after reconstruction. This effect can be evidenced by simulating intentional misalignment (not shown).

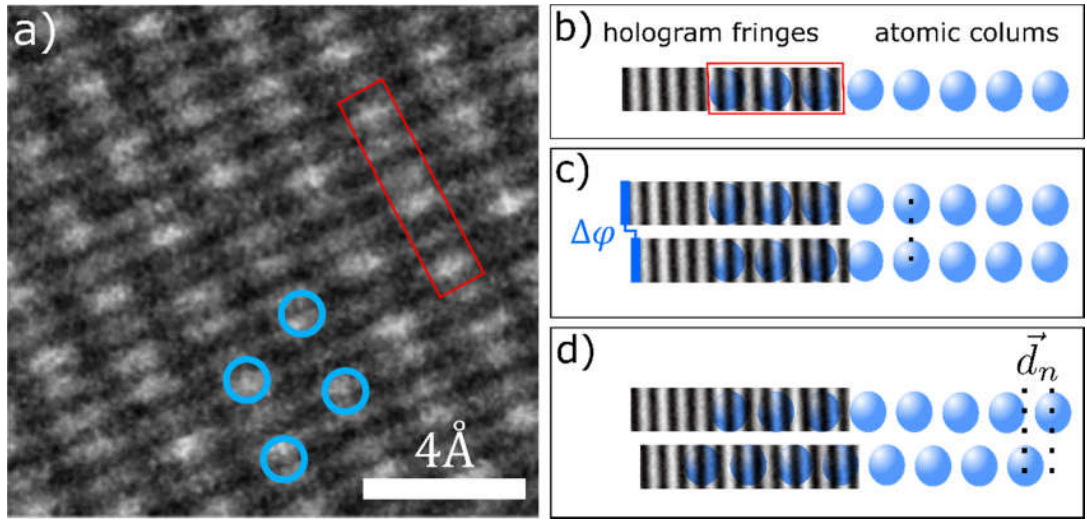

**Figure 2:** Schematic representation of undesired contrast changes from specimen drifts in holograms. **a)** Single frame (exposure time 1s) of a phase-shifting hologram series of pure platinum in [110] zone axis. Atomic column positions of one unit cell are marked with blue circles and are not easily identifiable by eye. **b)** Superposition scheme of fringes and atomic columns sketching the signal in the red box of a). **c)** Fringe shift $\Delta\varphi$ for any two holograms induced by a beam tilt, highlighted in blue. **d)** Effect of an additional specimen drift $\vec{d}_n$ in the two holograms. Note that the fringe distortion due to the local phase shift from the atomic potentials has been omitted in the schematic drawings for clarity.

Specimen drift correction for conventional high-resolution TEM images typically involves cross-correlation procedures [39,40]. The latter cannot be applied to images of an electron hologram series in a straightforward way due to the presence of interference fringes and their shift relative to the sample. Figure 2 illustrates the problem at hand with part of a hologram (a) and a schematic drawing (b), showing atomic column contrast and interference fringes. In the absence of specimen drift, Figure 2 (c) shows the effect of phase shifts resulting from beam tilts within the holograms series and possible instabilities of the biprism. Finally, Figure 2 (d) shows the combined effect of phase shifts and specimen drift.

## 3.2 Specimen drift correction scheme for atomic-resolution holograms

A straight forward approach would use standard drift correction techniques on filtered images obtained by application of a centerband aperture to each hologram in order to select the Bragg spots which resemble the Fourier transformed (FT) HRTEM image (Figure 3). A more advanced form of this concept has been used successfully for the averaging of conventional off-axis holograms in a setup, where the Fresnel streak of the Fourier transformed hologram is eliminated by use of two biprisms [41]. If only a single biprism is used, the Fresnel modulations cannot be eliminated in the same way. For the medium biprism voltages of 200-300 V used, which are preferable for increased visibility of the hologram fringes and thus the signal to noise ratio of the reconstruction, the Fresnel modulation of the hologram fringes hinders a precise drift correction at atomic resolution. In cross-correlation calculations, the presence of a Fresnel streak in the Fourier image of the hologram (marked red in Figure 3) extends into the centerband and thus leads to drift corrections with insufficient accuracy, since the cross-correlation will be weighted towards the shift of the interference fringes. Hence, a reduction of Fresnel streak contributions to the centerband is mandatory. Filtering in Fourier space by e.g. a Gaussian line filter clearly suppresses the Fresnel streaks, but frequently introduces artefacts resulting in residual drift correction errors of atomic positions.

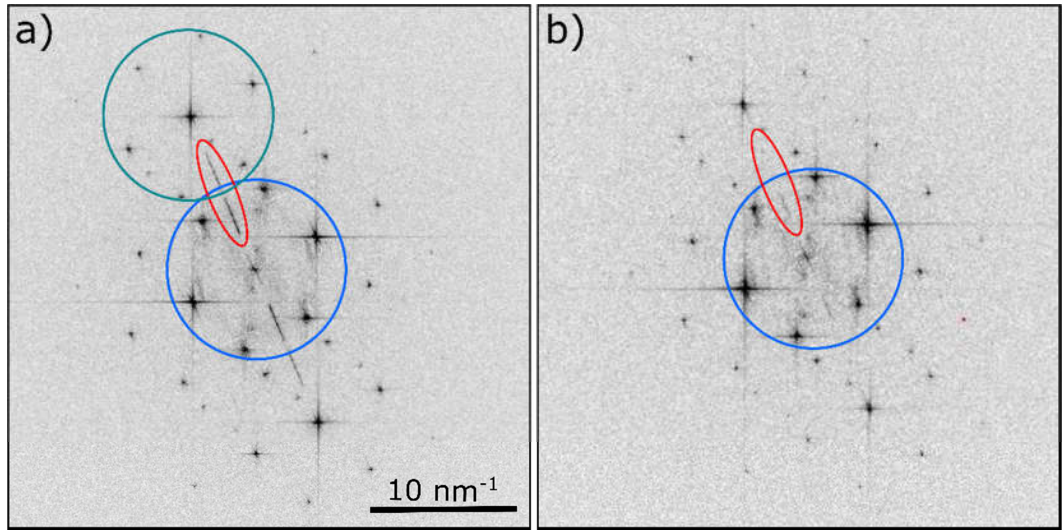

**Figure 3:** a) Fourier transform of a single hologram from a phase-shifting series on Pt [110]. The biprism voltage of 250 V separates the first order reflections of the centerband (dark blue) and the sideband (cyan). The overlap of the Fresnel streak (marked red) and the centerband impedes the application of the center band cross correlation method for drift determination. b) FT of a hologram from the series after division by an average reference hologram aligned as described in the text. The Fresnel streak is suppressed and the drift can be determined from the centerband cross correlation of different holograms of the series.

Instead, a robust and sufficiently accurate real space algorithm is proposed here, that requires the presence of an interface to the vacuum area in the hologram series. Each hologram is divided by an average reference hologram $I_{n,ref}$ resulting in an increased weight of Bragg peaks compared to the Fresnel streak. In order to apply this approach of drift correction, the biprism voltage must be selected as a trade-off between good visibility of interference fringes and a sufficient separation of centerband and sideband as shown in Figure 3 (a) and (b).

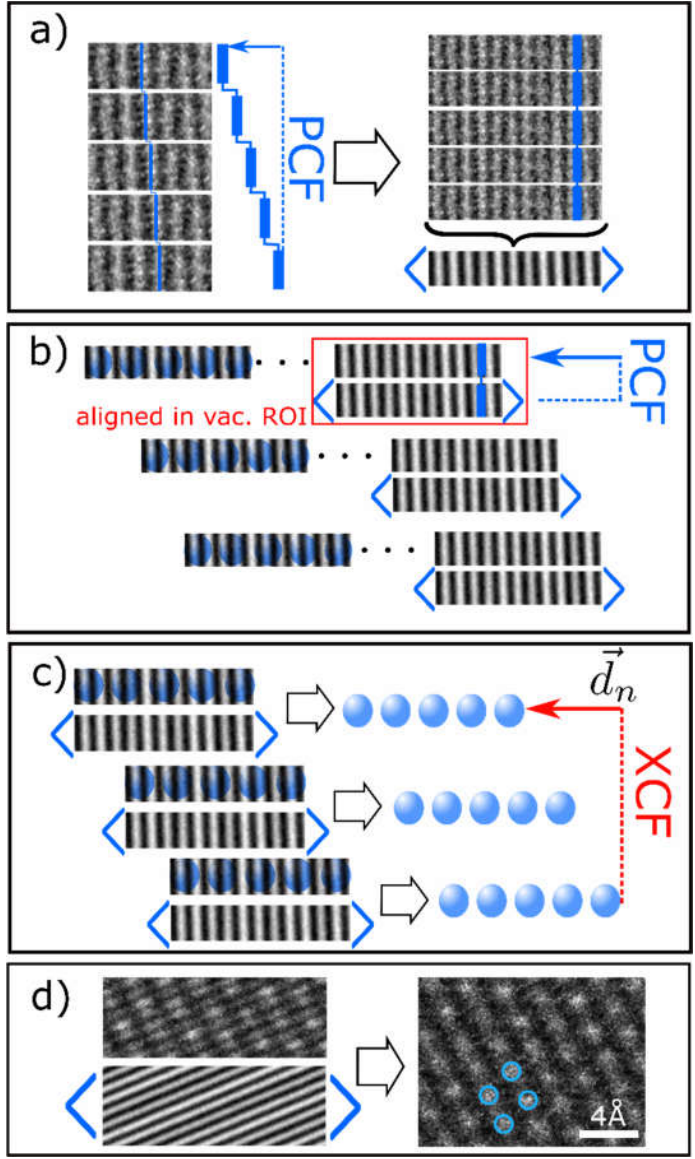

**Figure 4:** Scheme to extract the specimen drift vector $\bar{d}$ from the phase-shifted hologram series as described in the text. Atomic columns are depicted as blue spheres, fringe shifts between holograms are also highlighted in blue. **a)** Alignment of the reference PS-EH series for averaging using the phase correlation function (PCF). **b)** PCF alignment of the average reference hologram to each individual hologram of the PS-EH series. A region of interest (ROI) of the vacuum region was chosen for the procedure in both images. **c)** Division of PS-EH series and average reference hologram to determine the specimen drift vector $\bar{d}$ by using the conventional cross correlation function (XCF) as described in the main text. **d)** Example of the division of a single frame of the PS-EH series by the averaged reference hologram. Atom positions of one unit cell are marked in the resulting image on the right-hand side.

Figure 4(a)-(d) provides a detailed illustration of the real space algorithm used to determine the specimen drift $d$. The essential steps can be summarized as follows:

1. As a first step (Figure 4 (a)), the reference hologram series $I_{n,ref}(x,y)$ is averaged to reduce the noise. Prior to averaging, of course, the holograms have to be aligned to revert the effect of the beam tilt and biprism drift. This is implemented using the phase correlation function (PCF) as introduced in [41,42], i.e.

$$PCF(r) = T^{-1}\left[\tilde{A}(q)\frac{\tilde{I}_n(q)\tilde{I}_{n-1}(q)}{\|\tilde{I}_n(q)\tilde{I}_{n-1}(q)\| + \varepsilon}\right] \qquad (9)$$

where $\tilde{A}$ represents an aperture function, and $\varepsilon$ a small constant to prevent singularities. It should be noted, that using the PCF instead of a conventional cross-correlation function (XFC) increases the weight of the Fresnel modulations compared to the carrier frequency. This weight change supports the accurate alignment of the Fresnel modulations when calculating $I_{n,ref}$.

2. In the second step (Figure 4 (b)) $I_{n,ref}$ is aligned to each hologram $I_n(x,y)$ of the series with the specimen in field of view. For this second alignment a region of interest in the vacuum region of both holograms is chosen and the corresponding shift vector is calculated again using a PCF.
3. In the last step (Figure 4 (c)), each hologram is divided by the aligned averaged reference hologram. This empirical method suppresses the effect of Fresnel fringes as shown in Figure 3 (b) andFigure 4 (d) for typical atomic resolution conditions with small fringe displacements. Among other attempts, e.g. subtraction or position selection, the division has turned out to be the most robust method.
4. The hologram image series is now prepared for centerband extraction by Fourier filtering. The choice of the filter is material specific and depends on the experimental conditions as well. In the present case a top-hat aperture around the centerband is used to exclude the sideband information. Finally, the specimen drift is determined by the maximum of the XCF of this filtered hologram series. The extracted sample drift $\overline{d_n}$ is then applied to the original raw data and the reconstruction is performed on this specimen-drift corrected series.

The general quality of the reconstruction process can be estimated once the images $a(x,y)$, $b(x,y)$ and $\Theta(x,y)$ have been calculated. For this purpose, the fit-predicted hologram series is calculated by using Eq. (2) with the $\varphi_n$ determined during the reconstruction process. This result is then compared to the experimental hologram series. Image wise comparison of the fit-predicted holograms with the experimental holograms allows for the calculation of a goodness of the fit-value for each reconstructed pixel (for details see SI section 3). The vacuum region can be used as benchmark level for the achievable optimum, that is not affected by sample drift. A drastic breakdown of this benchmark value in the specimen region indicates a failure of the drift correction.

With the drift vector $\overline{d_n}$ of $I_n(x,y)$ being determined, each raw-data hologram of the series $I_n(x,y)$ is shifted by the according vector components $I_n(x - d_{n,x}, y - d_{n,y})$. Note that this computational image displacement for specimen drift $\overline{d_n}$ correction alters the phase $\varphi_n$. It is therefore crucial for the reconstruction procedure to determine $\varphi_n$ as described by Eq. (4) only after the specimen drift correction.

Each individual pixel of CCD and CMOS cameras does not show exactly the same response to the input signal because of imperfections, e.g. of gain correction. These artefacts are obviously not affected by any specimen drift. Consequently, they are averaged by performing the fit, if specimen drift is absent, e.g. in the empty reference series $I_{n,ref}(x,y)$. In contrast, in $I_n(x,y)$ the computational image displacements that are necessary to correct the specimen drift will lead to a reconstruction based on different pixel responses. This is leading to specific camera artifacts within the reconstruction of $I_n(x - d_{n,x}, y - d_{n,y})$.

Applying the same drift correction shift vector also to the raw empty reference stack $I_{n,ref}(x,y)$ before the reconstruction enables using the same combination of pixels for the sample and the reference reconstruction. Thus, it allows compensating these camera artefacts during the empty reference correction. Note that the initial phase shift $\varphi_{n,ref}$ of the reference hologram series must be updated after drift correction. After the specimen drift correction, the reconstruction of both series is done according to Section 2.2.

# 4 Experimental details

In order to demonstrate the high spatial frequency sensitivity at Angstrom resolution of drift-corrected PS-EH, we have chosen platinum in [110] zone axis as a model system. A platinum foil has been rolled to ≈ 50 μm thickness, a circular disc stamped out, and milled with low-energy argon ions ($E_{min}$ = 300 V) in a precision ion polisher to less than 10 nm edge thickness.

The thickness has been measured in energy-filtered imaging mode (EFTEM) with a Gatan Quantum GIF using the log-ratio technique for a large field of view. The mean free path for platinum was $\lambda_{Pt}$ = 67.7 nm according to [43] for 300 kV and 45 mrad collection angle.

A FEI Titan 80-300 G2 ETEM at 1.4 Mx magnification operated at 300 kV and equipped with a Gatan Ultrascan 1000 XP CCD was used to record PS-EH data.

The PS-EH series, including the reference, are generated with a biprism in the selected area aperture port at a voltage of 250 V leading to a mean fringe contrast of about 17% [13%, 19%]. The change of the visibility is mainly related to the biprism stability. It is not corrected during reconstruction due to the presence of the Fresnel modulations (see SI section 4). An unreconstructed reference hologram and an estimation of the Fresnel signal ratio is shown in the SI section 5.

The recorded holograms cover a field of view of ~14 x 14 nm with 2048 x 2048 pixels. The fringe spacing is 14 px and the pixel spacing 0.068 Å/px. The total hologram width has a size of ≈ 51 nm, so that a part could be selected that possess minimal distortions from Fresnel fringes and vignetting [44].

For platinum [110], 250 V was sufficient to separate the first order Bragg spots of the centerband from those of the sidebands, which enable us to select the centerband by the drift correction aperture while providing the highest visibility.

The beam tilt direction was chosen perpendicular to the biprism filament, which in turn was oriented perpendicular to the biprism holder axis in order to minimize mechanical instability. The tilt was calibrated with a wobbling procedure to tilt magnitudes of 0 to 4.5 μrad corresponding to $\varphi_n \approx [0, 4\pi]$ (see SI section 1). Each PS-EH series consists of 51 holograms acquired with comparatively short exposure times of one second to allow for *a posteriori* drift correction between the holograms. The empty reference PS-EH series is acquired under identical illumination conditions directly subsequent. The image shift deflectors are used to move the field of view, the stage remaining the same position to prevent the mechanical instabilities. A lower limit estimate for the dose rate was $30245\ e^-\ nm^{-2}\ s^{-1}$ as calculated from the reference series of 51 slices.

In order to provide an additional validation of the determined amplitude and phase from phase shifting holography, an additional defocus series has been taken at a close-by specimen position at 1.1 Mx and the relative defocus was varied from -30 nm to +30 nm in 3 nm steps. The results of these measurements are provided in the supplementary information.

Simulated wave functions for Pt-[110] are calculated by multi-slice technique using Dr Probe [45] software. The modulation transfer function of the CCD, temporal coherence envelope according to $d_{info} = 0.1$ nm measured for our instrument following [46], and thermal diffuse scattering by frozen lattice simulations with the room temperature Debye-Waller factor, expressed by the isotropic B-factor $B_{iso,Pt}$=0.384 $\left[\text{Å}^2\right]$ [47], are included in the simulation.

While the instrument is equipped with an image aberration corrector (CEOS CETCOR), the residual aberrations were also numerically corrected using a minimum amplitude contrast criterion approach [48]. The first order lens aberrations are found by varying defocus and twofold astigmatism simultaneously. Initial values of $C_1$ are chosen in a window of $\pm 3$ nm width around the minimum edge contrast defocus of 21 nm in 24 steps. The twofold astigmatism was varied from 0 nm to 50 nm in one nanometer steps and the according angle from 0° to 360° in one degree steps. For $C_3$, -1 µm was chosen fixed, as estimated by the CEOS corrector software. A second-order correction did not significantly improve the results. The resulting phase plate can be found in SI section 6.

## 5 Results

The PS-EH reconstruction is demonstrated by acquiring a beam tilt series of holograms at high magnification. The material is well-described in literature with a face-centered cubic (Fm-3m) unit cell. Platinum will be of further interest for environmental studies of surface potentials due to its electrocatalytic activity.

Figure 5 shows the reconstructed image-wave, split into amplitude and phase, obtained from the PS-EH study on platinum in [110] zone axis, after the drift correction as described in section 3 has been applied. The experimental defocus was later determined to be 22 nm (overfocus). The Fourier image of the phase (Figure 5 (c)) reveals coherent Bragg intensity from {-333} spots at distances > 10 nm$^{-1}$. The studied sample region contains some inhomogeneities in regards to local thickness, mistilt and surface relaxation. The surface reveals a tendency towards faceting, while evidence of preparation-induced damage (amorphization) is minimal apart from point defects. In order to reduce statistical noise, we choose to average 5x5 supercells of equal size from a homogeneous sample region, as shown in the insets of Figure 5 (a) and (b). This avoids the broad blurring effect that band-pass filtering would produce, provided that all supercells are perfectly identical and aligned. This is of course an approximation despite possibly existing inhomogeneities.

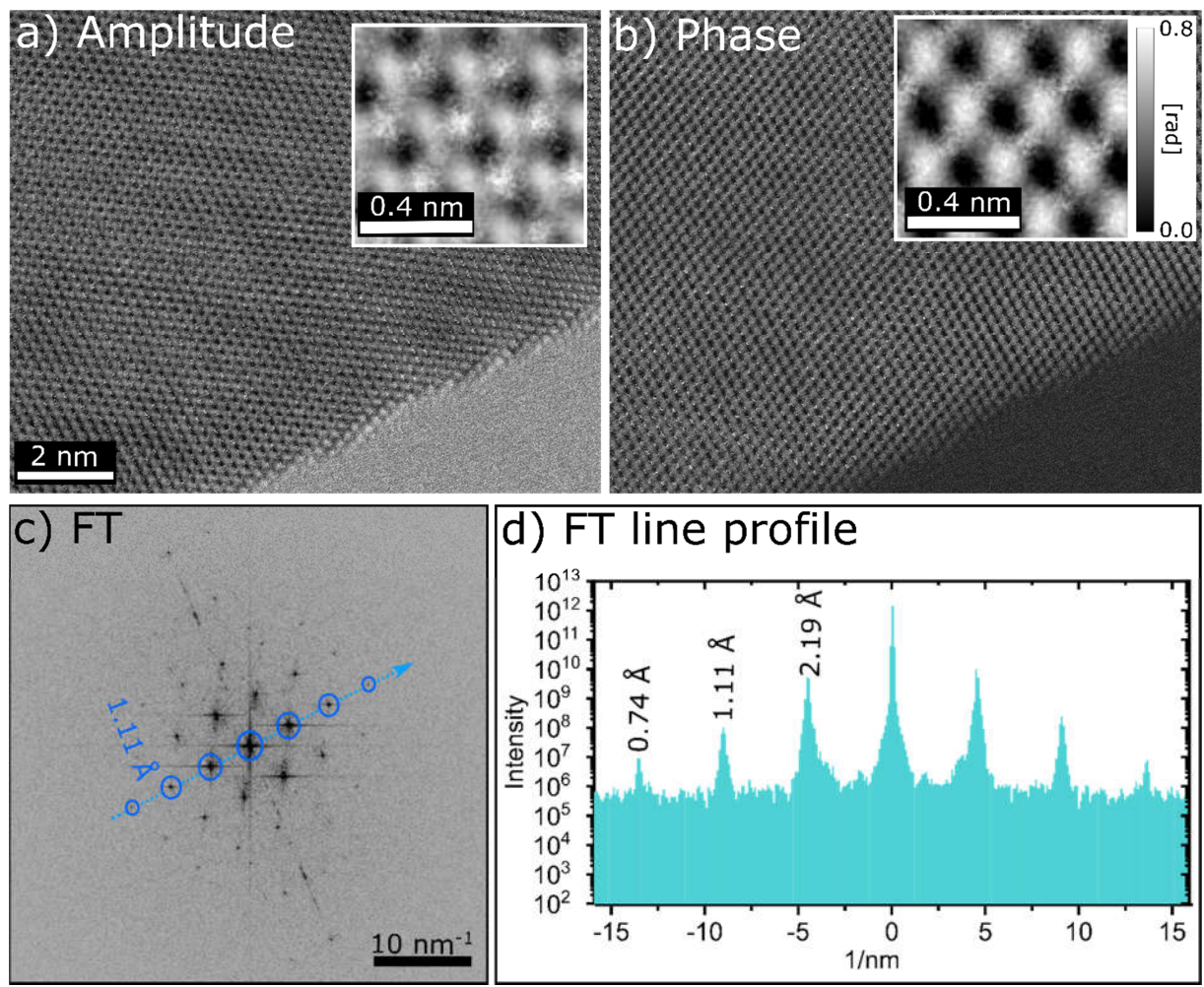

**Figure 5:** a) Amplitude of the PS-EH image-wave and b) corresponding reconstructed phase. c) Logarithmic modulus of the complex signal FT shown in a)-b). d) Line profile of the amplitude of c) along indicated direction. The Bragg spots up to third order are visible in the FT. The insets in a) and b) show the lateral average of 5x5 supercells from a homogeneous region of the sample.

For a perfect reconstruction without specimen drift and Fresnel distortions, amplitude and phase information would be completely separated. In this case, the spatial frequency in Figure 5 (d) would indicate an achieved spatial resolution of the phase information of up to 0.74 Å. However, the presence of residual signals of the fringe carrier frequency and the Fresnel streak in the Fourier image needs to be taken into account. They are an indication of imperfections in fitting which could lead to an intermixing of amplitude and phase signals. In the Fourier image in Figure 5 (c), the residual signal of Fresnel streak and carrier frequency are below 2% and 1% of their original value, respectively (see SI section 7).

Since the hologram stack was aligned by correcting the drift of the underlying object, Fresnel modulations from the biprism edge will be shifted across the series according to the sample drift vector and the biprism drift. Consequently, intensity variations deviating from the ideal cosine fit may serve as an indicator of the expected residual error.

In order to exclude the intermixing of amplitude and phase signals and to validate the obtained wave function, we chose to compare the reconstruction with multislice simulations. For best match between simulation model and experiment, the effect of residual objective lens aberrations must be corrected. The numerical phase plate ($C_1$=22.25 nm, $A_1$= (9 nm, 310°), see SI section 4 ) for first-order correction was determined from the exit wave by the minimum amplitude contrast criterion [48] restricted to defocus and twofold astigmatism. Since the image corrector is well-tuned before the experiment (A2/B2 < 50 nm, C3 < 2 μm, A3 and S3 < 1 μm), an attempt to correct residual higher-order aberrations did not produce a significant improvement in the amplitude contrast criterion, in particular since effects of residual sample tilt may be mistaken for lens distortions.

As starting point for the multislice simulations, an approximate measure for the sample thickness was obtained from the log-ratio technique as described in the experimental details section. A large field of view including the PS-EH measurement region is shown in the SI section 8. The platinum edge thickness at the PS-EH position, assuming a mean free path of 67.7 nm in Pt, is measured to approximate 2.5 nm with a standard deviation of $\pm 1.0$ nm. For the multi-slice image simulations, different sample thicknesses up to 15 times the (110) lattice plane distance, corresponding to 8 nm, were used, due to the uncertainty of the log-ratio technique for such small thicknesses. Simulated exit-wave super cells were extracted after N slices, corresponding to different simulated thicknesses. For each thickness the phase of one simulated supercell has been spatially averaged at zero defocus and compared to the experimental one. The best match between model thickness and average experimental phase is found at 2 nm, which agrees with the log-ratio results.

Figures 6 and 7 show the comparison between multislice simulation and propagation of the experimental wave on absolute intensity scale for the amplitude and Figure 7phase signals, respectively. Simulation and experiment are compared in terms of symmetry of the 5x5 supercell-images. A false colour scale was chosen to enhance the contrast for the eye. To compare the absolute values of amplitude and phase, exemplary line profiles along the dashed lines are additionally shown. The experimental line profile is depicted in blue, the simulation in red. The absolute values are scaled to the magnitude of the reference wave (amplitude) and relative to the vacuum phase shift (phase). Note that the average signal in the shown supercells does not vary with the defocus. The change of the defocus only redistributes signal in regions outside the line profile.

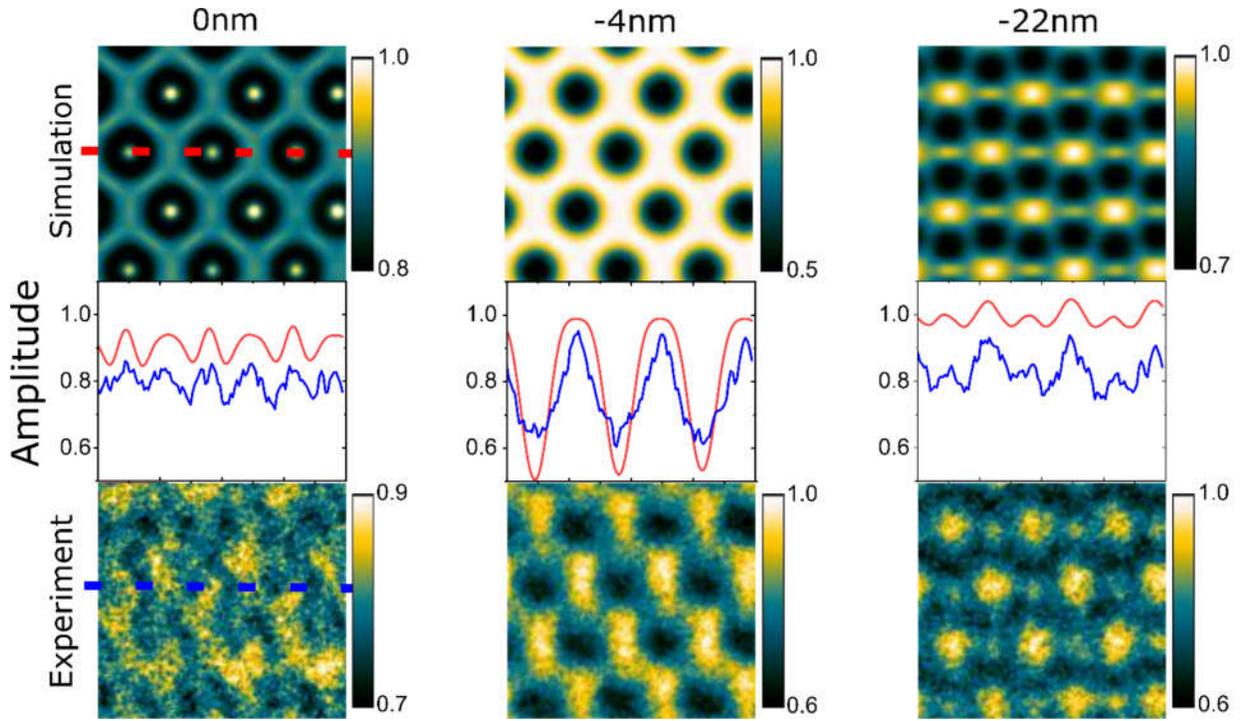

**Figure 6:** Comparison of absolute amplitude between a multi-slice simulation at a thickness of 2 nm and a reconstructed, first order aberration corrected PS-EH image-wave of platinum. The simulated supercell (top) and the experimental averaged 5x5 supercell (bottom) are shown for minimum contrast defocus (0 nm), Lichte defocus (-4 nm) and -22 nm underfocus, where strong features are visible at a distance of 1.3 Å. Line profiles along the indicated dashed lines compare simulated (red) and experimental amplitude (blue). The total width of the supercell corresponds to 0.8 nm. Amplitude root mean square deviations between simulation and experiment are below 0.16.

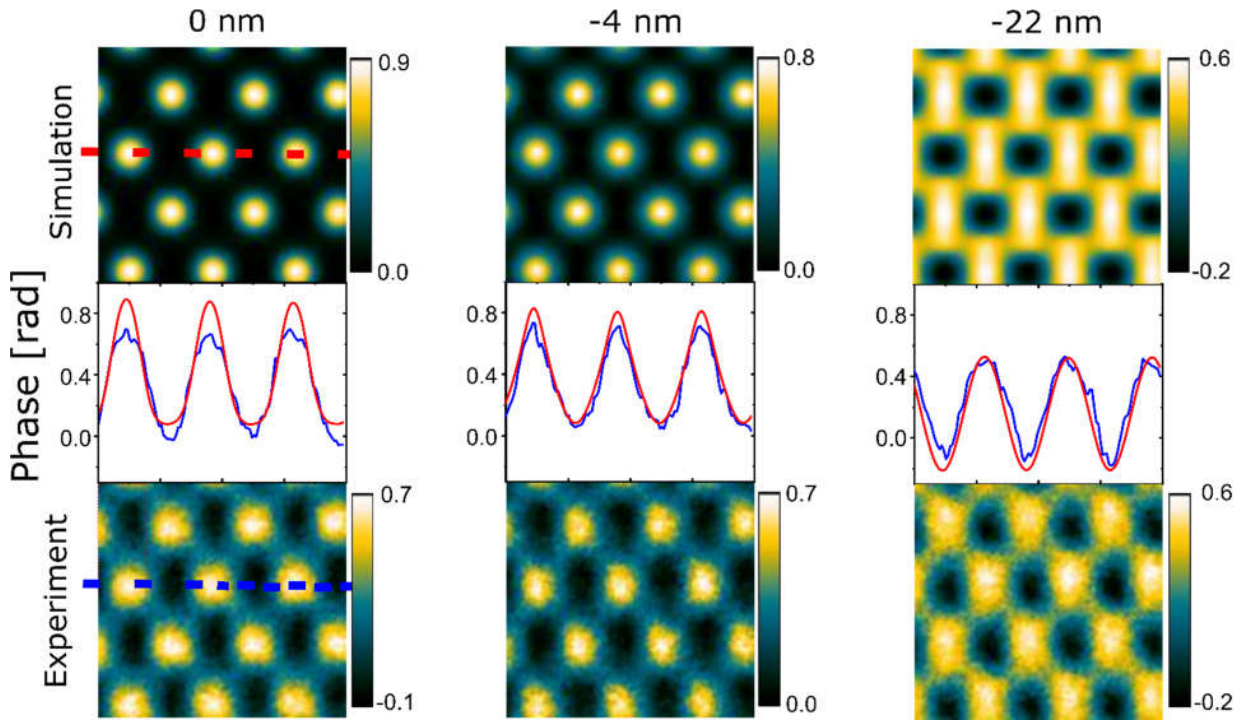

**Figure 7:** Comparison of the phase between a multi-slice simulation at a Pt thickness of 2 nm and a reconstructed, first order aberration corrected PS-EH image-wave of platinum. The simulated supercell (top) and the experimental averaged 5x5 supercell

(bottom) are shown for minimum contrast defocus (0 nm), Lichte defocus (-4 nm) and -22 nm underfocus. Line profiles along the indicated dashed lines compare simulated (red) and experimental phase (blue). The total width of the supercell corresponds to 0.8 nm. Phase deviations between simulation and experiment are below 0.11 rad root mean square difference.

Three particular points of interest are shown: The point of minimum amplitude contrast, which we define as the relative zero defocus in the experiment by the aberration fitting. The underfocused wave close to the Lichte defocus (-4 nm), where the phase is maximally localized [49]. And the wave at large underfocus (-22 nm), where amplitude contrast appears at spatial frequencies of 1.3 Å that approach the information limit.

# 6  Discussion

The simulation shows a high degree of agreement with the reconstructed experimental wave function. The expected symmetry in amplitude and phase images was found for a broad defocus range with only small deviations. The comparison on the absolute scale shows root mean square deviations below 0.16 (relative to the reference wave magnitude) for the amplitudes and below 0.11 rad for the phase. Inelastic processes which are not captured in the simulations, and other experimental inaccuracies, likely result in a systematic error which is in an acceptable range. In particular, at very small sample thicknesses, residual drift correction errors, local mistilt and small deviations from the nominal thickness, averaged over the supercells, can have a significant influence on amplitude contrast and mean amplitude, respectively.

We decided to neglect the effect of asymmetric distortion of the modulation transfer function (MTF) in the experimental and simulated waves that are shown in Figs. 6 and 7. A realistic handling of the asymmetry would require either MTF deconvolution of single holograms, resulting in amplification of high-frequency noise. Alternatively, it would need a direct integration of the MTF effect in the PS-EH reconstruction formalism, which is beyond the scope of this study. Instead, a symmetric MTF is applied to the simulated wave to account for information loss due to the camera system. Nevertheless, the effect of this formally improper MTF modelling was simulated in the noise-free case (see SI 9). The maximum resulting phase distortion was $2\pi/76$, which nearly matches the achieved accuracy of one pixel's phase within standard deviation in the reconstruction.

As an additional validity check, an experimental defocus series has been taken and its amplitude compared to the amplitude of the defocus-propagated wave function (see SI section 10). This reduces the possible error from aberration fitting, since both waves suffer from nearly identical aberrations. The amplitudes are in good agreement and show a similar break of the diagonal next neighbour symmetry as in Figure 6 (-4 nm), which may indicate residual specimen tilt. This is also suggested by the trend of further multislice simulations (SI section 11). However, the comparison of amplitude contrast between defocus series and PS-EH is not extracted from exactly the same sample position and therefore thickness and tilt do not necessarily agree.

Further disentangling of specimen tilt and higher-order aberrations would require a higher precision of thickness determination. It has been previously reported, that the absolute phase shift at small sample thicknesses of several atoms cannot be linearly extrapolated with thickness due to dynamic interactions which are particularly important for heavy atoms [23]. Our multislice simulations show that dynamic scattering behaviour becomes relevant in Pt after a few atoms, so that the average phase does not increase monotonically with thickness. This behaviour leads to an uncertainty in the thickness determination from phase shift alone of up to 2 times the (110) lattice plane distance, i.e. 1 nm, for specimens below 5 nm thickness. For further increasing sample thicknesses, tilt can result in non-linear behaviour of the reconstructed phase [50] and the contrast changes may thus be misinterpreted as higher-order aberrations. In a strict sense, thickness, tilt and aberrations have to be determined concurrently, which is beyond the scope of this work.

In summary, the multislice comparison demonstrates that amplitude and phase are well separated and therefore spatial resolution indeed approaches the information limit of our microscope of about 0.8-1.0 Å, as suggested by the second and third order Bragg reflections in the phase image of Figure 5 (d). This is also supported by contrast features in the amplitude images at -22 nm defocus in Figure 6, where contrast peaks arise at a distance of 1.3 Å between the atom positions.

In addition to the spatial resolution, the phase sensitivity of the PS-EH reconstruction needs to be addressed. As already mentioned above, we refer to phase sensitivity as the standard deviation of the reconstructed phase in the vacuum region. It can be measured at the vacuum region of the phase reconstruction [51], without averaging, by calculating the standard deviation of line profiles perpendicular and parallel to the fringes. Using non-averaged line profiles with a width of one pixel (0.07 Å) yields a standard deviation of 0.087 rad and thus a phase sensitivity of $2\pi/84$ in parallel and $2\pi/74$ in perpendicular direction. This compares to 70-80% of the phase shift induced by the projected potential of a single isolated Pt atom. If the width of the line profiles is extended to 1 Å, corresponding to the information limit, $2\pi/181$ and $2\pi/171$ are achieved in parallel and perpendicular directions, respectively. The resulting increase of phase sensitivity by averaging is lower than expected for uncorrelated pixel noise alone. This can be attributed to the influence of systematic errors, e.g. the residual Fresnel- and carrier frequency signal visible in Figure 5 (c). Note that no filter has been used so far, so that these values provide a lower limit for the phase resolution. Possible approaches to improve phase resolution by eliminating frequency information above the information limit are e.g. the use of a low-pass filter or other nonlinear filtering methods (median-, wiener filter) to further smooth the reconstructed signals. Unlike regular EH reconstruction that is intrinsically low-pass filtered, PS-EH is subjected to pixel shot noise. By applying a low-pass filter on the phase map with a spatial frequency above the achieved resolution, namely 1 Å, the phase sensitivity becomes $2\pi/452$. For the low pass filtering a square averaging kernel of 14 x 14 pixels was used and the standard deviation of a line scan of one pixel was evaluated. This way the achieved phase sensitivity can be compared to Fourier EH reconstructions.

# 7 Summary

Using the method of phase shifting electron holography in a Titan 80-300 kV aberration corrected environmental TEM equipped with a single biprism, we have demonstrated accurate electron wave phase and amplitude reconstruction at a spatial resolution close to the information limit of the microscope of ~ 1 Å. The resulting phase sensitivity at the information limit measures to about $2\pi/452$. The new key step is the implementation of a specimen drift correction scheme that allows to align the phase-shifted hologram series by well-established cross-correlation procedure. The main requirement is to supress or ideally eliminate Fresnel and residual hologram carrier frequency signals, in order to extract a sufficiently clean Bragg diffraction signal that can be used for drift correction.

The method is verified by comparing the reconstructed and corrected wave to frozen lattice multislice simulations, where an excellent match is obtained. The remaining small deviations between model and simulation parameters tend to amplify when carrying out a numerical propagation. However, this mainly affects the amplitude image which is slightly reduced in magnitude due to inaccuracies that are not fully captured in the simulation. On the other hand, an excellent match in the phase is revealed.

The developed software package for the phase shifting holography acquisition and analysis is made accessible in [36] and empowers non-dedicated holography microscopes with atomic resolution phase measurement capabilities. The method allows for resolving the phase of atomically sharp interfaces while at the same time preserving the low-frequency information of long-range fields, ideally suited e.g. for the study of interfaces with electric field contributions at different length scales.

## Acknowledgement

This work is funded by the Deutsche Forschungsgemeinschaft (DFG, German Research Foundation) 217133147/SFB 1073, projects C02 and Z02. Use of equipment in the "Collaborative Laboratory and User Facility for Electron Microscopy" (CLUE) at University of Goettingen is gratefully acknowledged. The authors would like to thank Juri Barthel from the Ernst Ruska-Centre Jülich for valuable discussions on frozen lattice multislice simulations.

## Literature:

[1] J.M. Cowley, Twenty forms of electron holography, Ultramicroscopy. 41 (1992) 335–348. https://doi.org/https://doi.org/10.1016/0304-3991(92)90213-4.

[2] J. Dietrich, D. Abou-Ras, S.S. Schmidt, T. Rissom, T. Unold, O. Cojocaru-Mirédin, T. Niermann, M. Lehmann, C.T. Koch, C. Boit, Origins of electrostatic potential wells at dislocations in polycrystalline Cu (In, Ga) Se2 thin films, J. Appl. Phys. 115 (2014) 103507.

[3] X. Xu, Y. Liu, J. Wang, D. Isheim, V.P. Dravid, C. Phatak, S.M. Haile, Variability and origins of grain boundary electric potential detected by electron holography and atom-probe tomography, Nat. Mater. 19 (2020) 887–893.

[4] P. Kruse, A. Rosenauer, D. Gerthsen, Determination of the mean inner potential in III--V semiconductors by electron holography, Ultramicroscopy. 96 (2003) 11–16.

[5] A. Auslender, M. Halabi, G. Levi, O. Diéguez, A. Kohn, Measuring the mean inner potential of Al2O3 sapphire using off-axis electron holography, Ultramicroscopy. 198 (2019) 18–25.

[6] A. Auslender, G. Levi, V. Ezersky, S. Gorfman, O. Diéguez, A. Kohn, Mean inner potential of graphite measured by electron holography: Probing charge distribution and orbital diamagnetic susceptibility, Carbon N. Y. 179 (2021) 288–298.

[7] V. Boureau, D. Cooper, Highly spatially resolved mapping of the piezoelectric potentials in InGaN quantum well structures by off-axis electron holography, J. Appl. Phys. 128 (2020) 155704.

[8] B. Haas, J.-L. Rouviere, V. Boureau, R. Berthier, D. Cooper, Direct comparison of off-axis

holography and differential phase contrast for the mapping of electric fields in semiconductors by transmission electron microscopy., Ultramicroscopy. 198 (2019) 58–72.

[9] H. Lichte, P. Formanek, A. Lenk, M. Linck, C. Matzeck, M. Lehmann, P. Simon, Electron Holography: Applications to Materials Questions, Annu. Rev. Mater. Res. 37 (2007) 539–588. https://doi.org/10.1146/annurev.matsci.37.052506.084232.

[10] A. Tonomura, The quantum world unveiled by electron waves, World Scientific, 1998.

[11] R.E. Dunin-Borkowski, M.R. McCartney, B. Kardynal, D.J. Smith, Magnetic interactions within patterned cobalt nanostructures using off-axis electron holography, J. Appl. Phys. 84 (1998) 374–378.

[12] M.J. Hÿtch, F. Houdellier, F. Hüe, E. Snoeck, Dark-field electron holography for the measurement of geometric phase, Ultramicroscopy. 111 (2011) 1328–1337.

[13] C. Ophus, H.I. Rasool, M. Linck, A. Zettl, J. Ciston, Automatic software correction of residual aberrations in reconstructed HRTEM exit waves of crystalline samples, Adv. Struct. Chem. Imaging. 2 (2016) 1–10.

[14] F. Kern, M. Linck, D. Wolf, N. Alem, H. Arora, S. Gemming, A. Erbe, A. Zettl, B. Büchner, A. Lubk, Autocorrected off-axis holography of two-dimensional materials, Phys. Rev. Res. 2 (2020) 43360.

[15] Special Issue: Electron Interference Microscopy, Microscopy. 70 (2021). https://academic.oup.com/jmicro/issue/70/1#1211063-6125690.

[16] T. Tanji, K. Urata, K. Ishizuka, Q. Ru, A. Tonomura, Observation of atomic surface potential by electron holography, Ultramicroscopy. 49 (1993) 259–264.

[17] G. Lole, V. Roddatis, U. Ross, M. Risch, T. Meyer, L. Rump, J. Geppert, G. Wartner, P. Blöchl, C. Jooss, Dynamic observation of manganese adatom mobility at perovskite oxide catalyst interfaces with water, Commun. Mater. 1 (2020) 1–10.

[18] H. Yoshida, Y. Kuwauchi, J.R. Jinschek, K. Sun, S. Tanaka, M. Kohyama, S. Shimada, M. Haruta, S. Takeda, Visualizing gas molecules interacting with supported nanoparticulate catalysts at reaction conditions, Science (80-. ). 335 (2012) 317–319.

[19] I.E.L. Stephens, A.S. Bondarenko, U. Grønbjerg, J. Rossmeisl, I. Chorkendorff, Understanding the electrocatalysis of oxygen reduction on platinum and its alloys, Energy Environ. Sci. 5 (2012) 6744–6762. https://doi.org/10.1039/C2EE03590A.

[20] P.A. Midgley, An introduction to off-axis electron holography, Micron. 32 (2001) 167–184.

[21] H. Lichte, M. Lehmann, Electron holography basics and applications, Reports Prog. Phys. 71 (2007) 16102.

[22] K. Ishizuka, Optimized sampling schemes for off-axis holography, Ultramicroscopy. 52 (1993) 1–5.

[23] M. Linck, B. Freitag, S. Kujawa, M. Lehmann, T. Niermann, State of the art in atomic resolution off-axis electron holography, Ultramicroscopy. 116 (2012) 13–23.

[24] E. Völkl, L.F. Allard, D.C. Joy, Introduction to electron holography, Springer Science \& Business

Media, 1999.

[25] A. Lubk, K. Vogel, D. Wolf, J. Krehl, F. Röder, L. Clark, G. Guzzinati, J. Verbeeck, Fundamentals of focal series inline electron holography, in: Adv. Imaging Electron Phys., Elsevier, 2016: pp. 105–147.

[26] W.M.J. Coene, A. Thust, M.O. De Beeck, D. Van Dyck, Maximum-likelihood method for focus-variation image reconstruction in high resolution transmission electron microscopy, Ultramicroscopy. 64 (1996) 109–135.

[27] A. Thust, W.M.J. Coene, M.O. De Beeck, D. Van Dyck, Focal-series reconstruction in HRTEM: Simulation studies on non-periodic objects, Ultramicroscopy. 64 (1996) 211–230.

[28] T. Niermann, M. Lehmann, Holographic focal series: differences between inline and off-axis electron holography at atomic resolution, J. Phys. D. Appl. Phys. 49 (2016) 194002.

[29] A. Lubk, F. Röder, Phase-space foundations of electron holography, Phys. Rev. A. 92 (2015) 33844.

[30] C. Ozsoy-Keskinbora, C.B. Boothroyd, R.E. Dunin-Borkowski, P.A. Van Aken, C.T. Koch, Hybridization approach to in-line and off-axis (electron) holography for superior resolution and phase sensitivity, Sci. Rep. 4 (2014) 1–10.

[31] C. Ozsoy-Keskinbora, C.B. Boothroyd, R.E. Dunin-Borkowski, P.A. van Aken, C.T. Koch, Mapping the electrostatic potential of Au nanoparticles using hybrid electron holography, Ultramicroscopy. 165 (2016) 8–14.

[32] V. V Volkov, M.G. Han, Y. Zhu, Double-resolution electron holography with simple Fourier transform of fringe-shifted holograms, Ultramicroscopy. 134 (2013) 175–184.

[33] V. Boureau, R. McLeod, B. Mayall, D. Cooper, Off-axis electron holography combining summation of hologram series with double-exposure phase-shifting: theory and application, Ultramicroscopy. 193 (2018) 52–63.

[34] Q. Ru, G. Lai, K. Aoyama, J. Endo, A. Tonomura, Principle and application of phase-shifting electron holography, Ultramicroscopy. 55 (1994) 209–220.

[35] K. Yamamoto, S. Anada, T. Sato, N. Yoshimoto, T. Hirayama, Phase-shifting electron holography for accurate measurement of potential distributions in organic and inorganic semiconductors, Microscopy. 70 (2020) 24–38. https://doi.org/10.1093/jmicro/dfaa061.

[36] J. Lindner, U. Ross, Developed Software @GitHub, (2023). https://github.com/SrcJonasLindner/phase-shifting-holography and https://data.goettingen-research-online.de/dataverse/phase-shifting-holography.

[37] D. Lei, K. Mitsuishi, M. Shimojo, M. Takeguchi, Reconstruction method for phase-shifting electron holography fitted with Fresnel diffraction affected fringes, in: Mater. Sci. Forum, 2015: pp. 215–221.

[38] R.A. McLeod, M. Kupsta, M. Malac, Determination of localized visibility in off-axis electron holography, Ultramicroscopy. 138 (2014) 4–12.

[39] M.J. Wester, D.J. Schodt, H. Mazloom-Farsibaf, M. Fazel, S. Pallikkuth, K.A. Lidke, Robust, fiducial-free drift correction for super-resolution imaging, Sci. Rep. 11 (2021) 23672.

[40] M. Kobayashi, A. Miyamoto, M. Tanaka, Y. Hoshino, A technique for high-precision drift correction in electron microscopes, Meas. Sci. Technol. 32 (2021) 125403.

[41] T. Niermann, M. Lehmann, Averaging scheme for atomic resolution off-axis electron holograms, Micron. 63 (2014) 28–34. https://doi.org/https://doi.org/10.1016/j.micron.2014.01.008.

[42] R.R. Meyer, A.I. Kirkland, W.O. Saxton, A new method for the determination of the wave aberration function for high resolution TEM: 1. Measurement of the symmetric aberrations, Ultramicroscopy. 92 (2002) 89–109.

[43] T. Malis, S.C. Cheng, R.F. Egerton, EELS log-ratio technique for specimen-thickness measurement in the TEM, J. Electron Microsc. Tech. 8 (1988) 193–200.

[44] H. Lichte, D. Geiger, A. Harscher, E. Heindl, M. Lehmann, D. Malamidis, A. Orchowski, W.-D. Rau, Artefacts in electron holography, Ultramicroscopy. 64 (1996) 67–77. https://doi.org/https://doi.org/10.1016/0304-3991(96)00018-6.

[45] J. Barthel, Dr. Probe: A software for high-resolution STEM image simulation, Ultramicroscopy. 193 (2018) 1–11. https://doi.org/https://doi.org/10.1016/j.ultramic.2018.06.003.

[46] J. Barthel, A. Thust, Quantification of the information limit of transmission electron microscopes, Phys. Rev. Lett. 101 (2008) 200801.

[47] H.X. Gao, L.-M. Peng, Parameterization of the temperature dependence of the Debye--Waller factors, Acta Crystallogr. Sect. A Found. Crystallogr. 55 (1999) 926–932.

[48] Z.K. Chen, W.Q. Ming, Y.T. He, R.H. Shen, G.S. Chen, M.J. Yin, J.H. Chen, Direct estimation and correction of residual aberrations in the reconstructed exit-wavefunction of a crystalline specimen, Micron. 157 (2022) 103247.

[49] H. Lichte, Optimum focus for taking electron holograms, Ultramicroscopy. 38 (1991) 13–22.

[50] F. Winkler, A.H. Tavabi, J. Barthel, M. Duchamp, E. Yucelen, S. Borghardt, B.E. Kardynal, R.E. Dunin-Borkowski, Quantitative measurement of mean inner potential and specimen thickness from high-resolution off-axis electron holograms of ultra-thin layered WSe2, Ultramicroscopy. 178 (2017) 38–47.

[51] T. Suzuki, S. Aizawa, T. Tanigaki, K. Ota, T. Matsuda, A. Tonomura, Improvement of the accuracy of phase observation by modification of phase-shifting electron holography, Ultramicroscopy. 118 (2012) 21–25.

# Supplementary Information

## Reconstruction of Angstrom resolution exit-waves by the application of drift-corrected phase-shifting off-axis electron holography

J. Lindner[1] and U. Ross[1], T. Meyer[1], V. Boureau[3], M. Seibt[4] and Ch. Jooss[1,2]

[1] Institute of Materials Physics, University of Goettingen, Friedrich-Hund-Platz 1, 37077 Goettingen

[2] International Center for Advanced Studies of Energy Conversion (ICASEC), University of Goettingen, D-37077 Goettingen, Germany

[3] Interdisciplinary Center for Electron Microscopy, École Polytechnique Fédérale de Lausanne, CH-1015 Lausanne, Switzerland

[4] 4th Institute of Physics – Solids and Nanostructures, University of Goettingen, Friedrich-Hund-Platz 1, 37077 Goettingen

## 1 Beam tilt calibration and typical zero disk shift

The remote control is done by C++ executables compiled for the Titan Scripting Package. Setter and getter for the beam tilt (gun tilt deflector or image rotation center) are implemented and accept command line arguments for the shift in [DAC] units.

And additional executable is used to implement the later described wobbling-calibration procedure.

All these executables are called by DM within the UI-Version to allow an automatized measurement of the tilt-magnitude calibration and the phase-shifting series.

The calibration that converts the tilt magnitude from the internal DAC-units to beam tilt angles in rad has been measured to 178 µrad by the shift of the zero disc for very high tilt amplitudes of $1 \cdot 10^{-2}$ [$DAC$] (Figure S1). This leads to tilt angles of < 4.5 µrad necessary for typical $4\pi$ phase shifts for 250 V biprism voltage. An initial phase measurement versus the hologram number including biprism drift can be found in Figure S2.

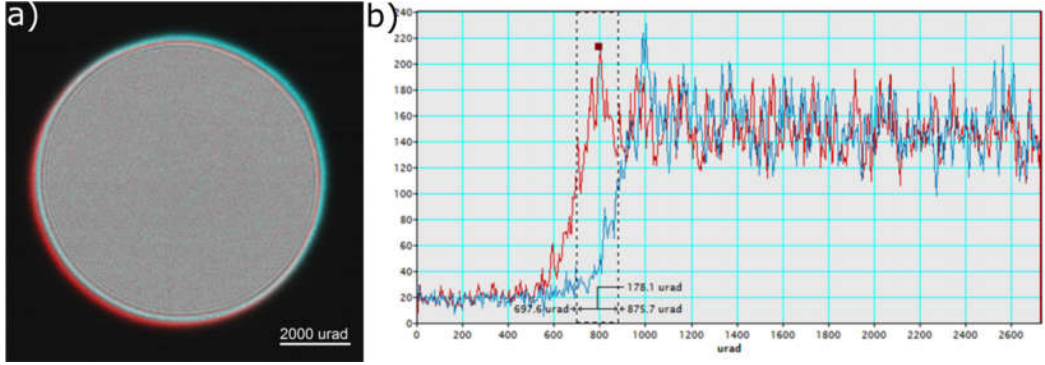

**Figure S1:** Determination of the gun tilt angle magnitude, disc size 4 mrad semi convergence angle. RGB image constructed from un-tilted zero disk (blue) and beam-tilted zero disk(red). The gun tilt excitation was 1 10$^{-2}$ [DAC] corresponding to 178.1 µrad. For typical a 250 V biprism tilt series covering [0, 4π] a tilt of up to 5 µrad was used.

For a given biprism voltage and a certain beam tilt angle the magnitude of the beam tilt in DAC-units have to be calibrated. To minimize the influence of mechanical vibration, the biprism should be oriented perpendicular to the holder axis. In the Titan the coordinate system of the beam deflection the y-direction is parallel to the holder axis. The wobbling calibration procedure takes a series of images. The exposure time of each image is synchronized with a tilt-ramp during this exposure. The number of points within the ramp n has to be chosen beforehand. Each image is acquired with a different maximal initial phase shift $\varphi_{max}$. The intensity of the acquired wobbling-image can be described as:

$$I_{wobbler}(\varphi_{max}, n) = \sum_{i=1}^{n} a(x,y) + b(x,y) \cos\left[\frac{2\pi x}{T_x} + \frac{2\pi y}{T_y} + \frac{\varphi_{max}}{(n-i)}\right] \quad (1)$$

If the standard derivation of each image is plotted against the $\varphi_{max}$ or tilt [DAC] the tilt magnitude needed to create a phase shift of $2\pi n$ can be measured by the distance of maxima of the standard derivative (compare Figure S3). The standard derivative is maximized if each step corresponds to $\frac{\varphi_{max}}{(n-i)} = 2\pi m$. This procedure allows a fast way of calibrating the needed tilt for arbitrary biprism orientations and angles and is included in the software package.

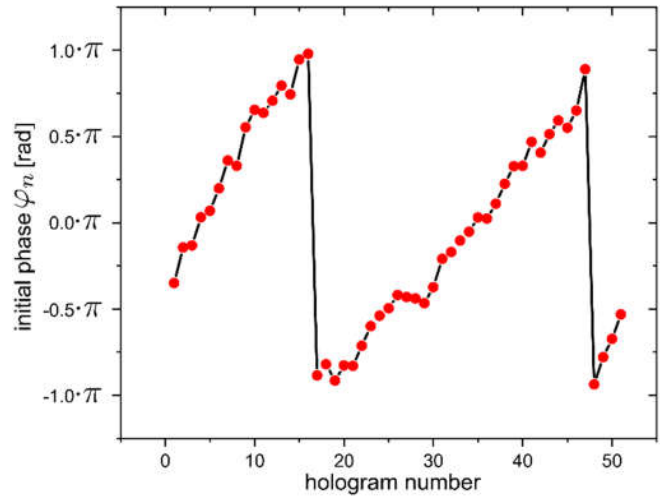

**Figure S2:** Initial phase shift $\varphi_n$ in the range of $[0, 4\pi]$ created by the gun tilt beam deflector and slightly distorted from the linear target ramp by biprism drift.

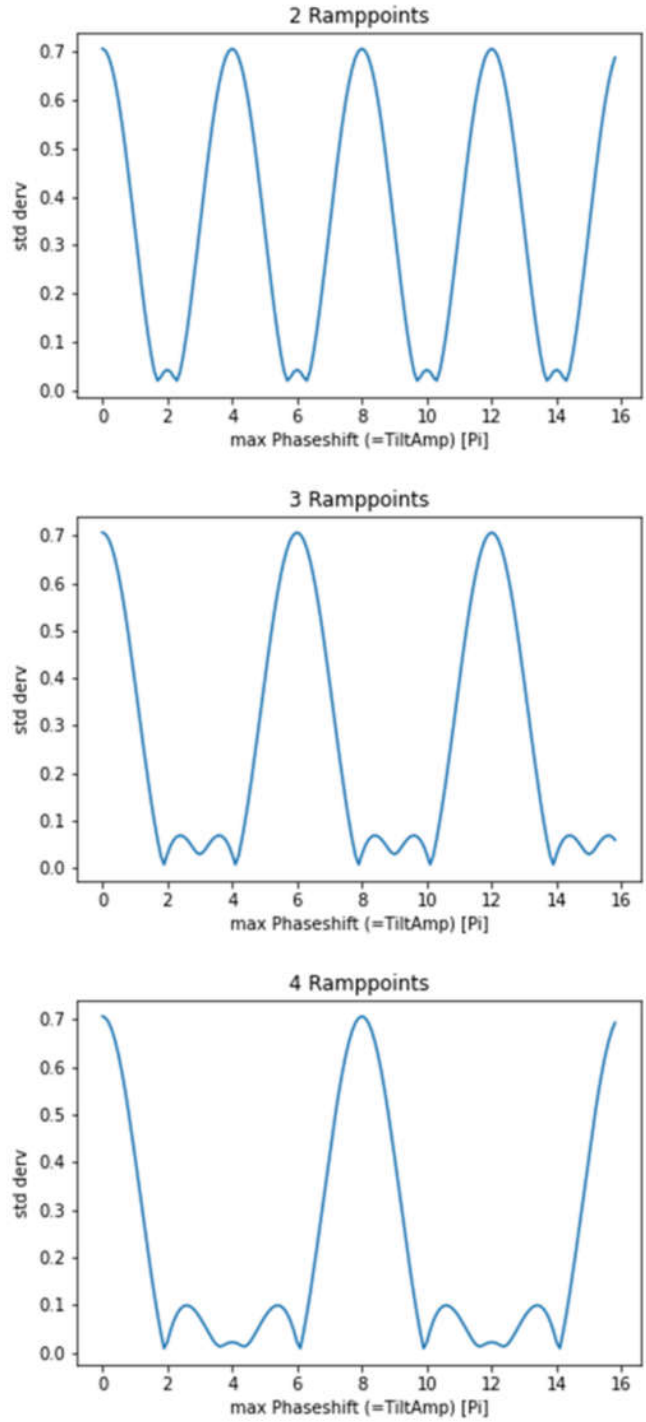

**Figure S3:** Standard derivative of $I_{wobbler}(\varphi_{max}, n)$ (Eq. (1)) for $n=[2,3,4]$ plotted vs $\varphi_{max}$. The distance of the maxima scales with $n \cdot 2\pi$, therefore the beam tilt amplitude for an arbitrary phase shift can be measured by the distance of maxima in the wobbling curve.

## 2 Fresnel diffraction fringe modulation

We have implemented the matrix reconstruction, that fits a Fourier series of finite order instead of a pure cosine function to the pixel-intensity data as suggested by Lei et al. [1].

$$I_n(x,y,n) = a(x,y) + \sum_j b^j(y) \; exp[\varphi_n^j + 2\pi i \, (q_c \quad q_j)] \qquad (2)$$

The Fourier components are subsequently numbered by j. The phase shift of the j-th component with the frequency $q_j$ is described by $\varphi_n^j$, while $q_c$ describes the hologram carrier frequency.

This should in principle increase the fit accuracy by giving account to the Fresnel modulations. We decided to not further follow this approach, because the gain in the goodness of fit was small (~ +3% $R^2$ gain) compared to the possible risk of overfitting. In our test we used integer fractions of the carrier frequency as Fourier series components, while it may be beneficial to find a proper way to extract them directly from the series or use a theoretical prediction by parametrisation of the Fresnel-modulation for the experimental conditions. Nevertheless, a Digital Micrograph script for this reconstruction is provided with the software.

## 3 Quantification of drift alignment and fitting quality

After the matrix reconstruction process the fitted parameters $a(x,y)$, $b(x,y)$ and $\Theta(x,y)$ can be used to build the intensity $\widehat{I_n}$ predicted by the least squares model for each hologram of the series by eq. (3) and the input phase values $\varphi_n$ used for the reconstruction.

$$I_n(x,y) = \underbrace{A_{ref}(x,y)^2 + A(x,y)^2 + I_{in}}_{a(x,y)} + \underbrace{2\mu(x,y) \; A_{ref}(x,y) \; A(x,y)}_{b(x,y)} \cos\underbrace{\left[2\pi(q_{cx} \; x + q_{cy} \; y) + \Phi(x,y)\right.}_{\Theta(x,y)} + \varphi_n\right] \qquad (3)$$

The goodness of fit $R^2$ can be calculated by

$$R^2 = 1 \quad \frac{\sum_{i=1}^{n}[I_i(r) \quad \hat{I}_i(r)]^2}{\sum_{i=1}^{n}[I_i(r) \quad \overline{I_i}(r)]^2} \qquad (4)$$

The according $R^2$-map is shown in Figure S4 (a). We observe increased values for the squared distance metric at the atomic columns of platinum compared to interatomic positions Figure S4 (b). The overall goodness of fit in the specimen region $R_{spec}^2$ is reduced compared to the vacuum region $R_{vac}$. A cosine fitting curve for the vacuum region is shown in Figure S4 (c) and a histogram of the goodness of fit map in Figure S4 (d). The $R^2$-metric can be used to get a rough measure of quality for the drift alignment procedure. If $R_{vac}^2$ roughly matches $R_{spec}^2$ the specimen drift was corrected successfully. From our experience a few percent difference is expected depending on the specimen and experimental conditions.

On the other hand a bad specimen drift alignment procedure shows drastic reduction in $R_{spec}^2$, since in this case the intensity values of the pixel projection belong not to the fitted cosine curve anymore. Due to noise, the Fresnel modulation of the fringes and residual drift alignment misfits, and the noncommutativity of pixels $R^2 \approx 1$ is not expected to be achieved in the used experimental setup.

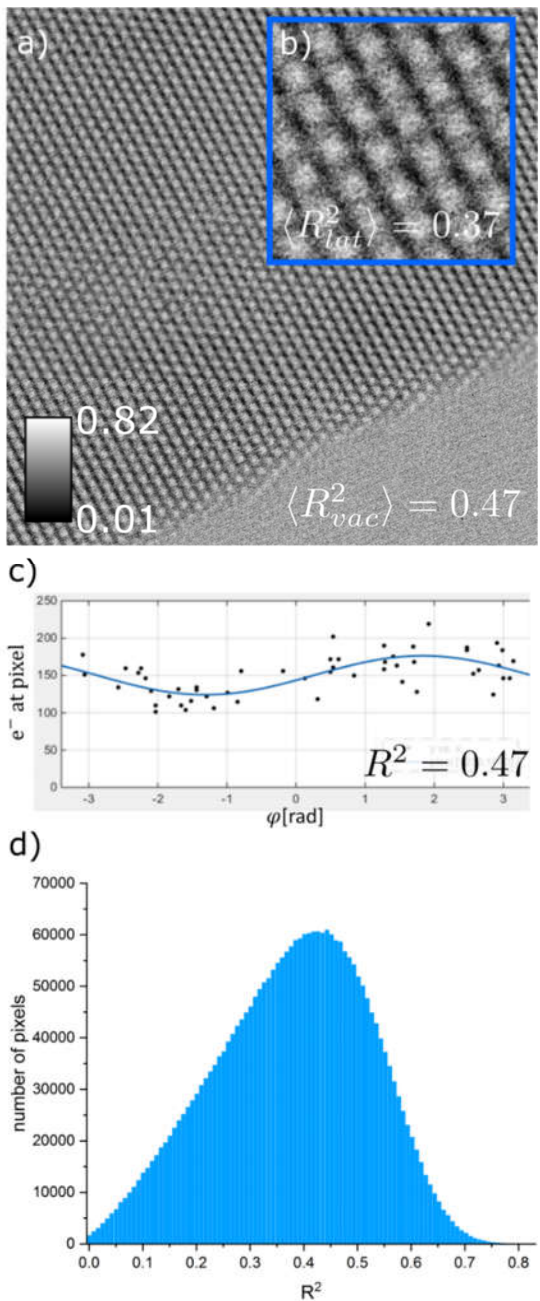

**Figure S4:** a) Map of the goodness of fit parameter $R^2$ that shows higher values at atomic positions and the average value for the vacuum. b) Inset is a zoom to an atomic lattice region. c) Example of a pixel projection cosine fit with corresponding to a $R^2 = 0.47$ d) Histogram of the $R^2$-values for all pixels shown in a).

## 4  Visibility changes

The visibility of the reference hologram series covers a range from 12.89% to 19.15% with a mean value of 16.98% (see Figure S5). These changes predominantly reflect the influence of the biprism stability.

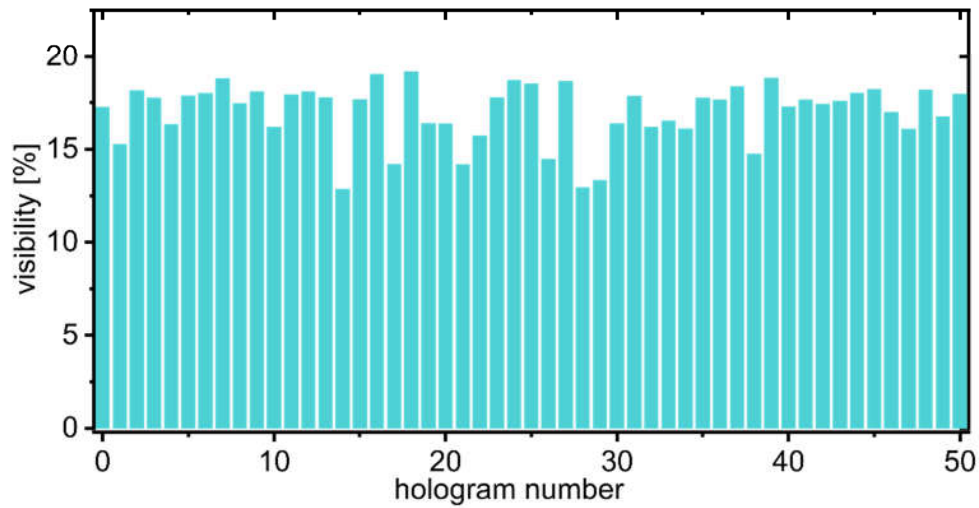

**Figure S5:** Variation of the global visibility in the reference hologram series.

The visibility variation of the series is not considered during the reconstruction due to the variation of Fresnel modulations across the field of view. A simple interpolation of the visibility cannot be carried out from the vacuum region above the sample, since the local visibility also changes, depending on which portion of the Fresnel modulation resides in the reference window.

# 5 Fresnel modulations

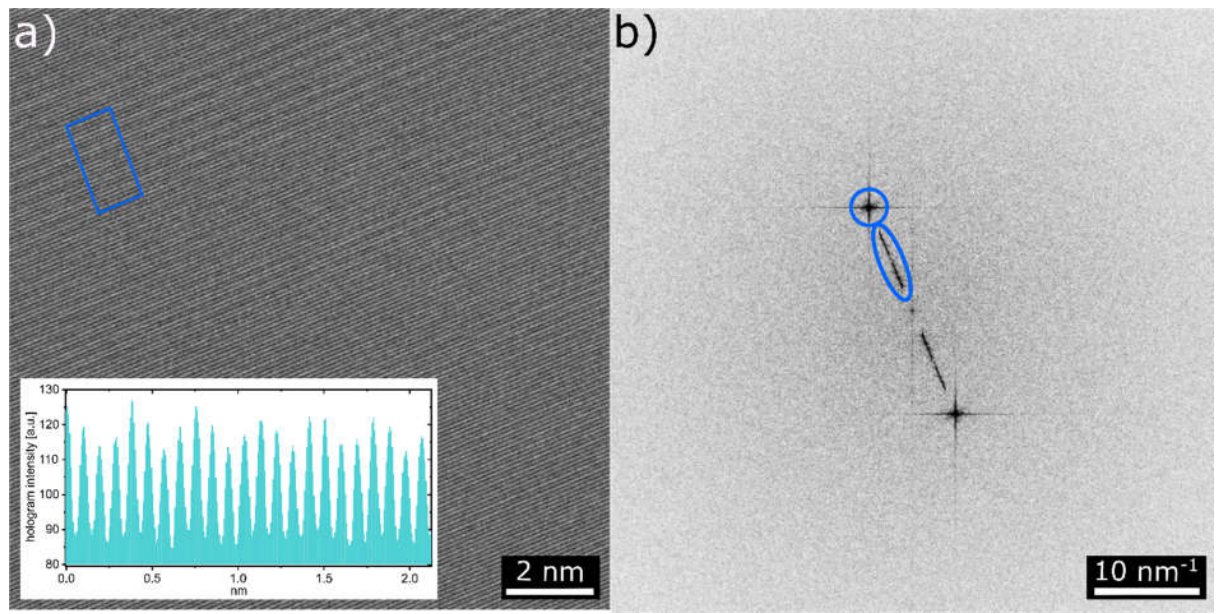

**Figure S6:** a) Hologram frame (1 sec exposure) from the raw reference series. The inset shows a line profile of a region that is strongly influenced by Fresnel signals (blue box). b) Corresponding modulus of the FT from a), showing the Fresnel streak and the sideband peak. The spectral weight of the Fresnel streak corresponds to 4.6% of the sideband.

# 6 Phase plate

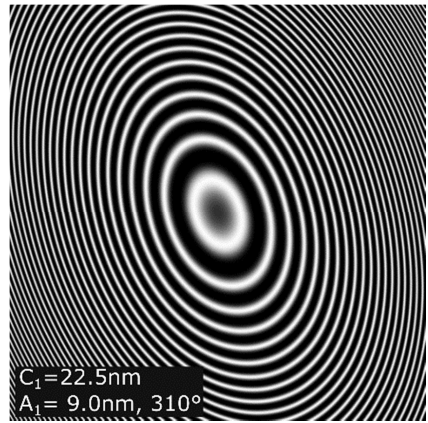

**Figure S7:** The numerical phase plate ($C_1$=22.25 nm, $A_1$=(9 nm, 310°)) resulting from the minimum amplitude contrast conditions.

# 7 Residual Fresnel signal

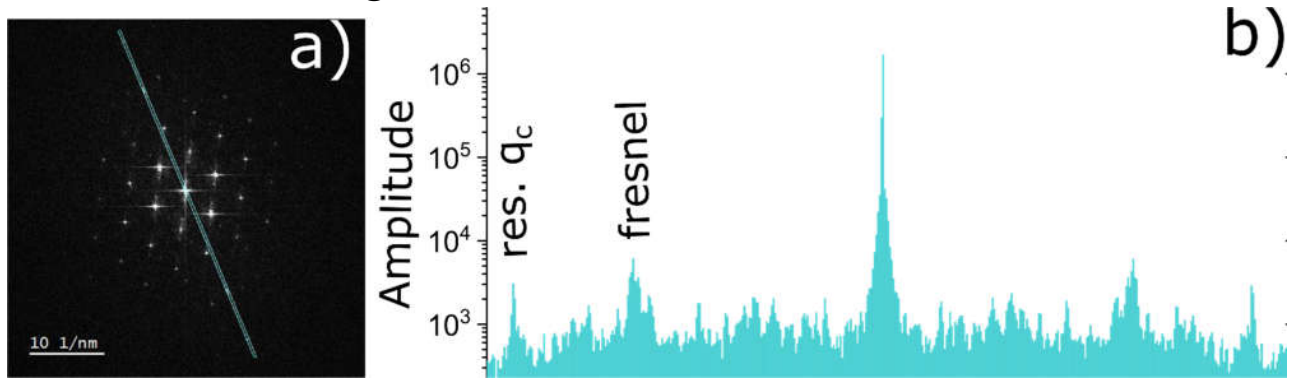

**Figure S8:** a) FTT of the reconstructed phase and b) line profile along the indicated area showing the residual Fresnel and fringe signal.

# 8 Thickness determination via log-ratio

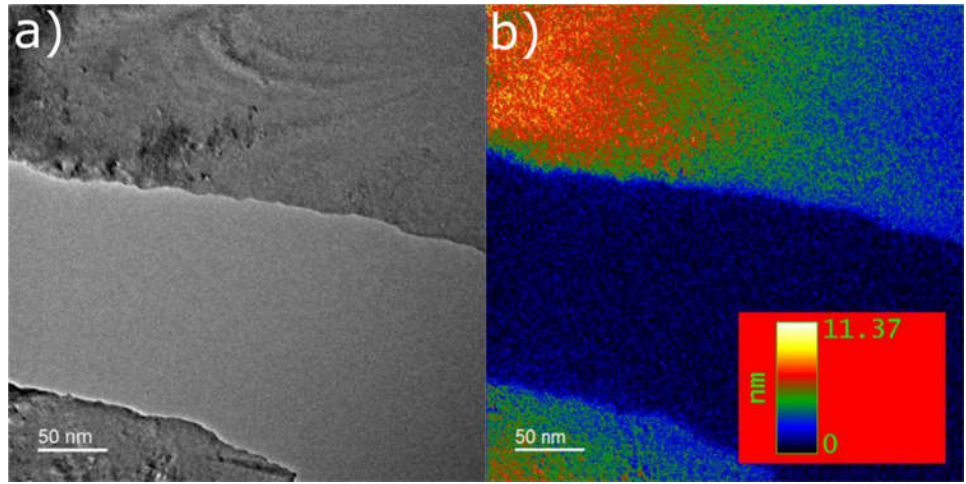

**Figure S9:** a) Overview image of the specimen edge. b) Thickness map from log-ratio technique using $\lambda_{Pt} = 67.7$ nm. The method evaluates the specimen thickness to approximate 2.5 nm at the brief PS-EH position

# 9 Handling of modulation transfer function (MTF)

If a hologram is recorded, the MTF of the camera leads to an asymmetric damping of the sideband signal due to the shifted reciprocal distance to the origin of the FT (compare Figure S10). This results in an additional phase distortion. To get rid of this distortion in the experimental data, low pass filtering and MTF deconvolution the raw data would be required. The low pass filter is needed because the high frequency part of the MTF will amplify the noise strongly. Finding the ideal filter to minimize the trade-off between filter artefacts, loss of information and noise amplification is beyond of the scope of this work.

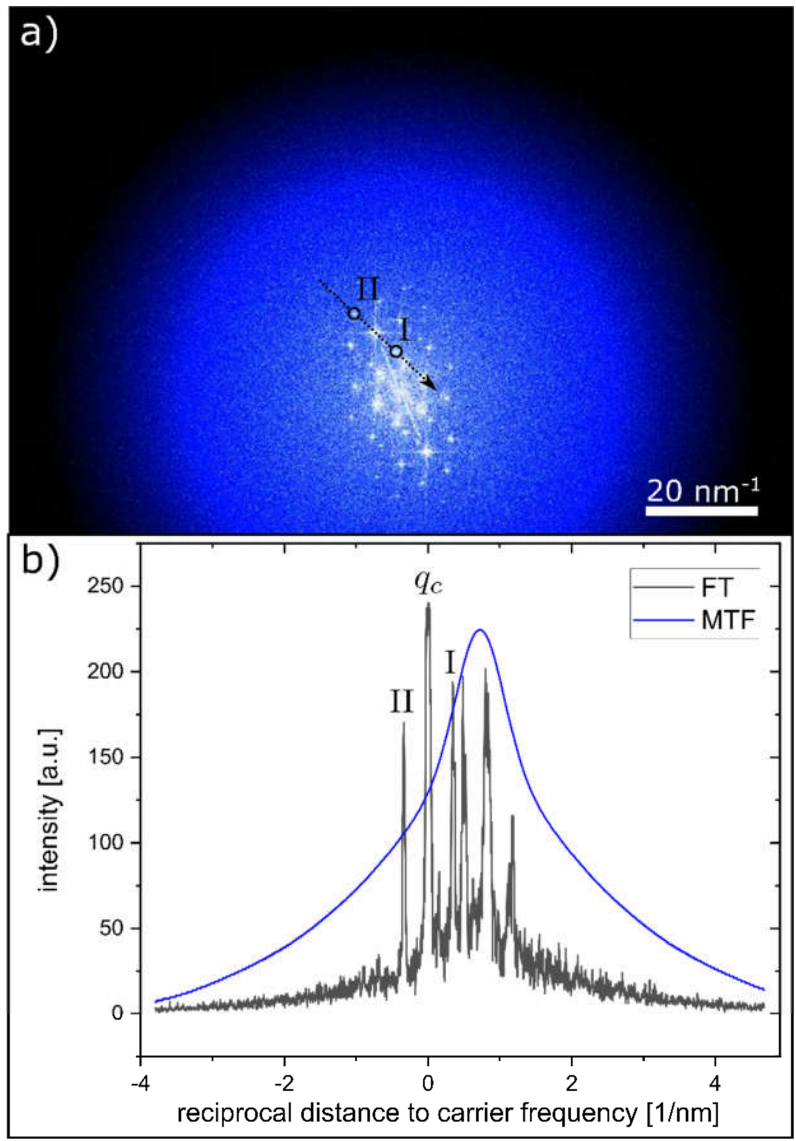

**Figure S10:** a) Overlay image of FT from a raw hologram (white) and camera MTF (blue). The roman numerals indicate two sideband peaks that are asymmetrically dampened by the MTF due to their different reciprocal distance from the FT origin. This creates a phase distortion in the reconstructed wave. b) Line profile along the arrow in a). Peak II exhibits a stronger dampening then peak I.

Instead, the distorting influence of the MTF has been simulated for the ideal, noise-free, case. For this purpose, the wave function from the multislice simulation without MTF was used to simulate an artificial hologram series. Every hologram was subsequently convoluted with the MTF and reconstructed as described in this work. The difference of the reconstructed phase with and without MTF convolution is shown in Figure S11. The simulation shows a maximum phase distortion of $2\pi/76$. This value approaches the achieved single pixel standard deviation of the phase in the experimental reconstruction.

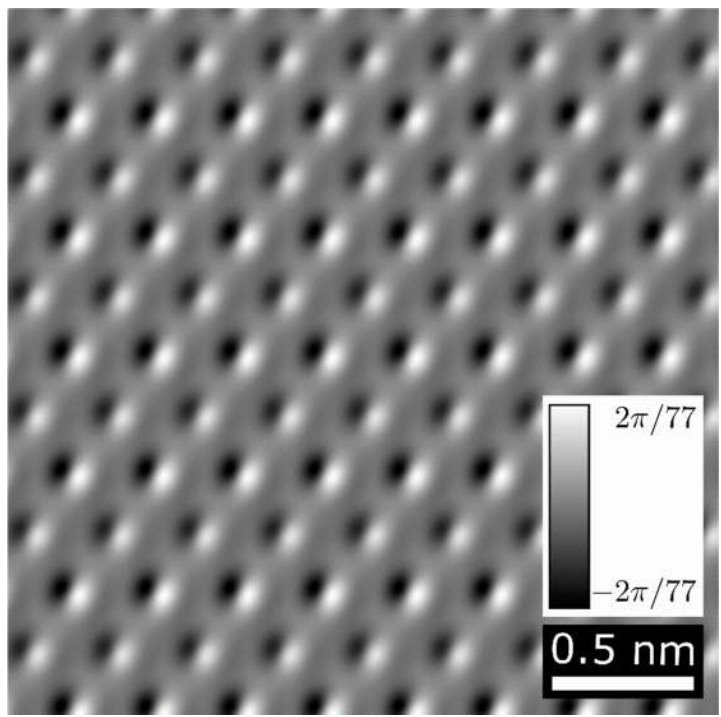

**Figure S11:** Phase distortion for the ideal noise-free simulation of the MTF influence. The simulated wave, described in the main text, was used to calculate an artificial hologram stack with and without MTF convolution before reconstruction. The phase distortion of the MTF was found to be at most $2\pi/76$, nearly matching the standard deviation of the reconstructed phase.

## 10 Experimental defocus series

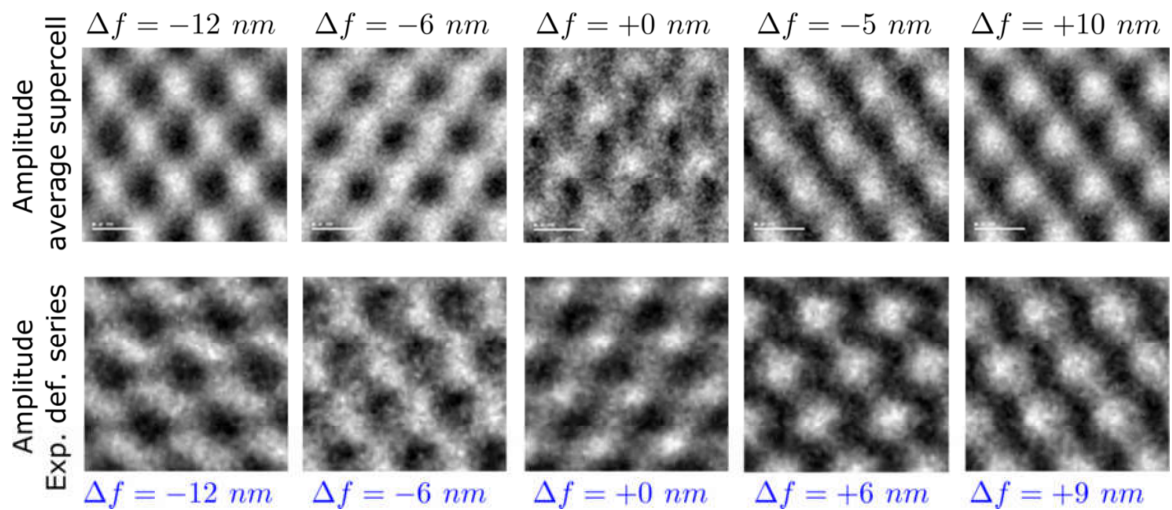

**Figure S12:** Comparison of amplitude defocus series to amplitude of reconstructed wave. Note that this defocus series was not captured at the exact same position as the phase shifting holography series.

## 11 Influence of simulated specimen tilt

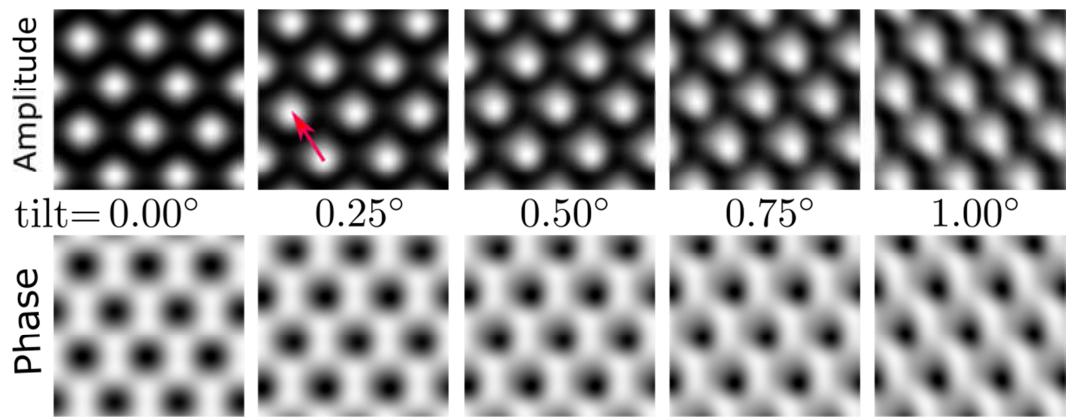

**Figure S13:** Influence of specimen tilt on amplitude and phase for a simulated thickness of 20 unit cells. The magnitude of the tilt is given between the images and a red arrow depicts the tilt direction.

## 12 Developed Software

Digital Micrograph scripts for the acquisition, beam tilt calibration and for the reconstruction of phase-shifting holography series and a user-documentation are provided at GitHub [2].

## Supporting Literature

[1]    D. Lei, K. Mitsuishi, M. Shimojo, M. Takeguchi, Reconstruction method for phase-shifting electron holography fitted with Fresnel diffraction affected fringes, in: Mater. Sci. Forum, 2015: pp. 215–221.

[2]    J. Lindner, U. Ross, Developed Software @GitHub, (2023). https://github.com/SrcJonasLindner/phase-shifting-holography and https://data.goettingen-research-online.de/dataverse/phase-shifting-holography.